\documentclass[useAMS,usenatbib]{mn2e}
\usepackage{graphicx}

\title[CoRoT\,102699796, the first metal-poor Herbig Ae pulsator: a
hybrid $\delta$ Sct-$\gamma$ Dor variable?]{CoRoT\,102699796, the first metal-poor Herbig Ae pulsator: 
a hybrid $\delta$ Sct-$\gamma$ Dor variable?\thanks{The CoRoT space mission, 
launched on December 27th 2006, has been developed and is operated 
by CNES, with the contribution of Austria, Belgium, Brazil , ESA (RSSD and Science Programme), 
Germany and  Spain.}\thanks{Based on observations collected at the
European Southern Observatory, Paranal, Chile. Proposal n. 082.D-0839A} }

\author[V. Ripepi et al.]{V. Ripepi$^{1}$\thanks{E-mail:
ripepi@oacn.inaf.it},
F. Cusano$^{1}$,
M. Di Criscienzo$^{2}$,
G. Catanzaro$^{3}$,
F. Palla$^{4}$, 
\and
M. Marconi$^{1}$,
 P. Ventura$^{2}$, 
C. Neiner$^{5}$, 
C. Catala$^{5}$,
S. Bernabei$^{6}$
\\
$^{1}$INAF-Osservatorio Astronomico di Capodimonte, I-80131, Napoli, Italy\\
$^{2}$INAF-Osservatorio Astronomico di Roma, I-00040, Roma, Italy\\ 
$^{3}$INAF-Osservatorio Astrofisico di Catania, I-95123, Catania, Italy\\
$^{4}$INAF-Osservatorio Astrofisico di Arcetri, I-50125, Firenze, Italy\\ 
$^{5}$LESIA, Observatoire de Paris, CNRS, UPMC, Universit\'e Paris Diderot, 92190 Meudon, France  \\
$^{6}$INAF-Osservatorio Astronomico di Bologna, I-40127, Bologna, Italy\\ 
}

\begin{document}

\date{}

\pagerange{\pageref{firstpage}--\pageref{lastpage}} \pubyear{2002}

\maketitle

\label{firstpage}

\begin{abstract}
We present the analysis of the time series observations of CoRoT\,102699796
obtained by the CoRoT satellite that show the presence of five independent
oscillation frequencies in the range 3.6-5 c/d. Using spectra acquired
with FLAMES@VLT, we derive the following stellar parameters:
spectral type F1V, T$_{\rm eff}$=7000$\pm$200 K, log(g)=$3.8\pm0.4$,
[M/H]=$-$1.1$\pm0.2$, $v$sin$i$=$50\pm5$ km/s, L/L$_{\odot}$=21$^{+21}_{-11}$.
Thus, for the first time we report the existence of a metal poor,
  intermediate-mass PMS pulsating star. Ground-based and satellite data are used to derive the
spectral energy distribution of CoRoT\,102699796 extending from the optical to
mid-infrared wavelengths. The SED shows a significant IR excess at
wavelengths greater than $\sim5 \mu$. We conclude that CoRoT\,102699796 is a young
Herbig Ae (F1Ve) star with a transitional disk, likely associated to the HII
region [FT96]213.1-2.2.

The pulsation frequencies have been interpreted in the light of the
non-radial pulsation theory, using the LOSC code in conjunction with
static and rotational evolutionary tracks.  A minimization algorithm
was used to find the best-fit model with M=1.84 M$_{\odot}$, T$_{\rm eff}$=6900 K which 
imply an isochronal age of t$\sim$2.5 Myr.
This result is based on the interpretation of the detected frequencies as $g$-modes of
low-moderate $n$-value. To our knowledge, this is the first time that such modes are
identified in a intermediate-mass PMS pulsating star. Since CoRoT\,102699796 lies in the region
of the HR diagram where the $\delta$ Sct and $\gamma$ Dor instability strips
intersect, we argue that the observed pulsation characteristics are
intermediate between these classes of variables, i.e. CoRoT\,102699796
is likely the first PMS hybrid $\gamma$ Dor-$\delta$ Sct pulsator ever
studied. 

\end{abstract}


\section{Introduction}



Asteroseismology of Herbig Ae stars offers a unique means to probe their
interior structure and to compare to evolutionary models. It is now well
established that these stars during contraction towards the main sequence
cross the instability strip of more evolved stars. These young, pulsating
intermediate-mass stars are collectively called PMS $\delta$ Sct  and their
variability is characterized by short periods ($\sim$30m$\div$5 h) and small
amplitudes \citep[from less than a millimag to a few hundredths of magnitude, see, e.g.]
[]{kurtz95,catala,ripepi06,zwintz08}.

The first theoretical investigation of the PMS instability strip based
on nonlinear convective hydrodynamical models was carried out by
\citet{marconi98}, who calculated its topology for the first three
radial modes. These authors also found that the interior structure of
PMS stars crossing the instability strip is significantly different
from that of more evolved Main Sequence stars (with the same mass and
temperature), even though the envelope structures are similar. The
subsequent theoretical work by \citet{suran01} made a comparative
study of the seismology of a 1.8 PMS and post-MS star. They
found that the unstable frequency range is roughly the same for PMS
and  post-MS stars, but that some non-radial ($g$) modes are very
sensitive to the deep internal structure. More recently,
\citet{griga06} produced a theoretical instability strip for PMS stars
for the first seven radial modes; \citet{ruoppo} derived a  model based
relation between the large frequency separation and the stellar
luminosity and effective temperature and developed a tool to compare
theory and observations in the echelle diagram. Finally
\citet{dicriscienzo2008} applied the ATON evolutionary code to the computation
of detailed grids of standard (non-rotating) and rotating pre-main
sequence (PMS) models and computed their adiabatic oscillation
spectra.

From the observational point of view, multi-site campaigns
\citep[e.g.,][]{ripepi03,bernabei09} and space observations with MOST
\citep[e.g.,][for NGC2264]{zwintz09} provided us with good sets of
frequencies to be compared with the models. 
In addition,
  spectroscopic studies \citep[see, e.g.][for the cases of HD104237
  and RS Cha, respectively]{fumel08,bohm09} demonstrated to be very
  valuable for line profile analysis and direct mode determination 
using tools like F2D \citep{kennelly94,kennelly96} or FAMIAS \citep{zima}.
However, the full
exploitation of the PMS $\delta$ Sct instability strip is still far
from being realized.  Dedicated observations of a large sample of PMS
stars covering the whole instability strip are needed to accomplish
this goal.  

In this context, a fundamental contribution can be provided by the data
coming from the space telescope CoRoT~\citep[Co\emph{nvection},
Ro\emph{tation and planetary} Tr\emph{ansits};][]{baglin07}. This 
satellite took images of the same region of the sky for about five consecutive months 
reaching relatively faint stars (up to
V$\sim$15 mag) in its``exo field'', with good precision
and very high duty cycle. Particularly interesting was the first
long-term pointing of CoRoT, the so-called ``Long Run a1'' ($LRa1$)
which imaged a region of the sky in the galactic anti-center where a
few distant Star Forming Regions (SFR) are present. 

With the aim of searching for possible PMS $\delta$ Sct variables present
in the CoRoT data, we have studied all the stars identified as $\delta$ Sct from the
CoRoT variable star classification group \citep{debosscher09}, looking
at near infrared (NIR) excess and emission in the Balmer lines, i.e. typical features of PMS objects.
We mainly used the 2MASS
photometry \citep{2mass} and the yet unpublished FLAMES@VLT spectroscopic survey of 
CoRoT variables in $LRa1$ performed by  C. Neiner and collaborators (ESO
proposal 082.D-0839A).
Among the about 76 stars classified as $\delta$ Sct with available
spectrum observed by CoRoT during $LRa1$, the only
candidate PMS $\delta$ Sct variable was CoRoT\,102699796, a
relatively faint star with V=15.51 mag, located in the direction of
Monoceros (RA,DEC 06:43:38.46 $-$01:07:46.7,  J2000). 
This paper is devoted to a complete characterization of this
star on the basis of time-series CoRoT observations and a variety of
additional data, such as VLT mid-resolution spectroscopy and  
optical-NIR ground-based, as well as space infrared photometry.  

In sections 2 and 3 we present the CoRoT photometry and the VLT 
spectroscopy, respectively. In section 4 we discuss the evolutionary
status of CoRoT 102699796 and its possible membership to the Star
Forming Region [FT96]213.1-2.2. In section 6 we interpret the observed 
oscillation frequencies to the light of non-radial pulsation theory; 
a summary of the major findings of present work closes this paper.

\begin{figure}
\centering
\resizebox{\hsize}{!}{\includegraphics[draft=false]{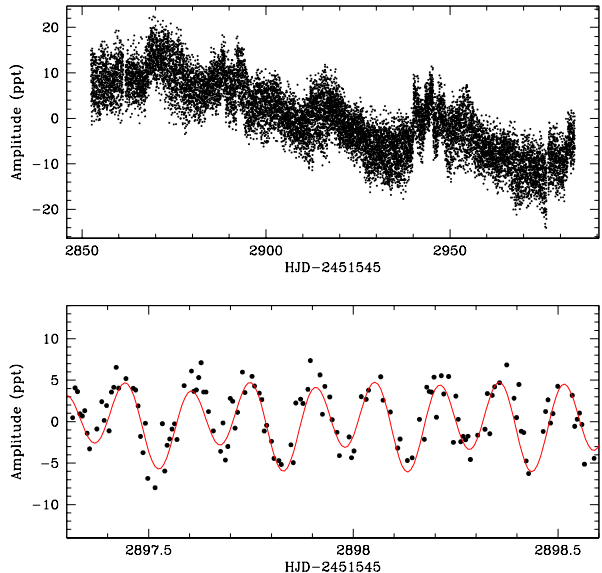}}
\caption{Upper panel: light curve of CoRoT\,10269979 before detrending
  (see text). Lower panel: selected portion of the light curve after
  detrending with overimposed the least-square fit (solid line) to the data obtained
  with the five ``genuine'' frequencies (see Sect.~\ref{comb}).}
\label{fig1}
\end{figure}

\section{Photometric observations with CoRoT}

CoRoT\,102699796 was observed by the CoRoT satellite in its ``exo
field'' during the Galactic anti-centre $LRa1$ run,
between October 27, 2007 and March 3, 2008. The data span about 131.5
consecutive days. The exposure time was of 512 seconds. We used the data corrected
to the N2 level \citep[the processing steps are described
in][]{auvergne09}, which contains 22036 flux measurements. The first
step of data processing consisted in removing all the points with
non-0 flag \citep[][]{auvergne09}, i.e. mainly removing those
measurements heavily affected by the transit through the South Atlantic
Anomaly (SAA). A few obvious outliers were rejected, too.  
The resulting light curve was transformed in part-per-thousand (ppt
hereafter) using
the formula 1000$\times$[(flux/$average$-1], where $average$ is the
mean over the whole light curve.
This procedure left us with 14997 useful data points which
are shown in Fig.~\ref{fig1} (upper panel). The light curve clearly
shows a monotonic dimming probably due to the ageing of the CCD and/or
optics of CoRoT, as well as 
shorter time scale variability, of the order of $\sim$15-30 d and
$\sim$3-5 d. The former ``periodicity'' could be ascribed to variable
dust obscuration due to the passing of circumstellar material 
in front of the star \citep[see, e.g.][]{vandenancker98}, whereas the
latter could be due to rotational modulation related to coronal activity \citep[see, e.g.][]{catala}.
However, given that the observed amplitudes are of 
only 1-2 hundredths of magnitudes for both types of variability, we
cannot exclude that all these long term variation are due to
instrumental effects.
Moreover, the light curve also 
appears affected by small``steps'', i.e. sudden jumps in
brightness of instrumental origin, common among the CoRoT light curves
in the ``exo field''
field \citep[see, e.g.][]{mislis10}. This occurrence makes it difficult to
understand whether a feature is intrinsic or due to instrumental
problems. 
To get rid of this effect, we decided
to apply a high-pass filter to the light curve, removing all the
long-term variability by filtering out all the frequencies lower than
0.50 c/d (all the periods larger than 2 days). This very simple
approach is effective in removing also the quoted ``steps'' because they are
not particularly strong in our case, and no specific treatment is
needed. 
We emphasize that the above procedures modify the variability features of the star
at frequencies much lower than those typical of the expected $\delta$ Sct
pulsation, therefore, they do not affect at all the frequency analysis which
is presented in the next section.

\subsection{Frequency analysis}

The pulsation frequency analysis was carried out with the $period04$
package \citep{lenz05}, which adopts both Fourier and least squares
algorithms, and permits the simultaneous fitting of multiple
sinusoidal variations, and thus does not rely on sequential
prewhitening. We have also used, for comparison purposes,  the  
SigSpec package, which follows a different approach \citep{reegen2007}. 

First, we calculated the spectral window (SW) of the data up to the
Nyquist frequency ($\sim$84 c/d), as shown in panel
a) of Fig.~\ref{fig2}. The SW shows several features due to the
satellite's orbital period of about
7000 s (variable up to four seconds), and its multiples. This is due to the removal of 
 data points during the passage through the SAA. Moreover, 
since the scattered light falling on the detector varies during the orbit of the
satellite, the periodogram of the star is expected to present the orbital period
($\sim$13.96 c/d) with the relative aliases and harmonics. Frequencies
close to these values will be rejected in the following analysis.

The Fourier transform of the data is shown in panel b) of
Fig. ~\ref{fig2}. It can be seen that there is a dominant frequency at
about 6.55 c/d and other signal up to the Nyquist frequency. Most of
the signal is represented by the aliases of the dominant frequency and
is removed by the prewhitening procedure. We extracted frequencies with $period04$ 
using as a limit for the last significant frequencies the widely used
empirical criteria of S/N=4 \citep[][]{breger93}. To estimate the
noise, following well established procedures \citep[see,
e.g.][]{rodriguez06,hernandez09}, we calculated the average amplitude of the residuals
  (after prewhitening all the significant peaks) in a frequency
  interval of width=5 c/d centred on the corresponding peak. 
Quantitatively the noise resulted to be quite flat 
beyond $\sim$1 (c/d) with an average value $\sim$0.03 (ppt). 
In this way we were able to extract 15
significant frequencies. Among these, four (at 27.9416, 41.91337
27.9379, and 13.9617 c/d) are related to the
orbital period of the satellite and will be ignored in the analysis. The
remaining eleven frequencies are listed in the first three columns of 
Table~\ref{tab1} together with the amplitude and S/N. The same 
procedure was carried out with $SigSpec$, obtaining the same results,
as shown in  Table~\ref{tab1} where $sig$ represent  the spectral
significance \citep[see][for an
explanation]{reegen2007}. Following \citet{reegen2007,
  kallinger08}, a value of  $sig$=5.46 
should be approximately equivalent to S/N=4.0. However, as shown
  in  Tab~\ref{tab1}, in our
  case  this S/N value corresponds to a $sig\sim$6.6. The origin 
of such a discrepancy, which was already found by other Authors in the literature, is not clear.  
A detailed discussion about this occurrence is beyond the scope of
present paper, indeed our results are not affected
at all, since our main discriminant for the significant frequencies
remains the S/N=4.0 criterion.
However, the interested reader can consult e.g. \citet{hernandez09} for a
discussion on this subject.

As a further check on the reliability of the frequency extraction, we
analysed separately selected portions of the time-series, with a size
of  about 10-15 d, and dealing with both ``detrended'' and raw
data (in this case avoiding portions including ``steps''). All
the relevant frequencies were found in all the cases, reinforcing our
confidence about the reliability of the results presented here.

The precision on the single frequencies is also given in
Table~\ref{tab1}. It has been estimated adopting the definition by 
\citet{kallinger08} : $\sigma f = 1/(T\sqrt{sig(A)}$, where $T$ is the
duration of the time-series and $sig(A)$ is the significance of the   
frequency with amplitude $A$ as calculated by $SigSpec$.We also report the uncertainty on the
frequencies on the basis of the Rayleigh criterion
(1/4T)\footnote{see \citet{kallinger08} and references therein for a discussion
  about this definition}=0.002
c/d. This more conservative estimate (at 
3-$\sigma$ level) is the value used for the
comparison to the theoretical models.

\subsection{Combination of frequencies}
\label{comb}
\begin{figure}
\centering
\resizebox{\hsize}{!}{\includegraphics[draft=false]{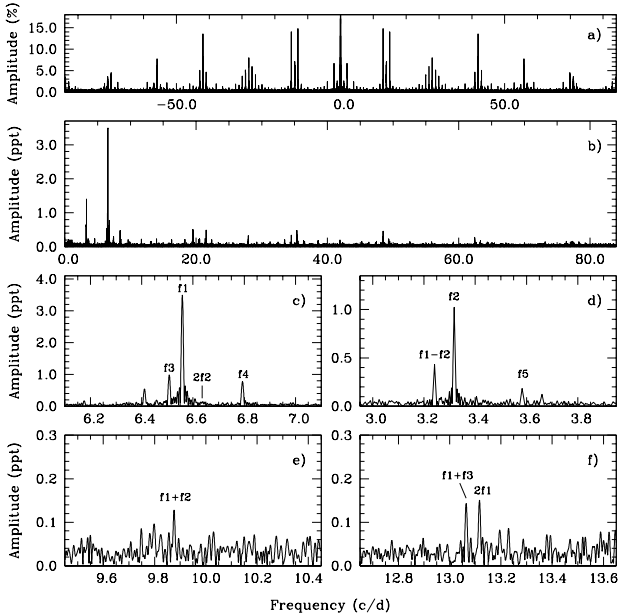}}
\caption{Panel a): spectral window of the data up to the
  Nyquist frequency. Panel b): Fourier Transform of the data. Panels
  c) to f): enlargements of Panel b) showing the relevant peaks
  extracted from the periodogram. Labels are the same as in Table~\ref{tab1}.}
\label{fig2}
\end{figure}

Linear combinations of the terms having the highest amplitudes
have been observed in low-amplitude $\delta$ Sct stars both from the
ground and from space  \citep[see, e.g.][]{breger05,poretti09}. 
To search for possible combinations among the frequencies observed 
in CoRoT\,102699796, we
decided to use the Combine package, an ad hoc computer program 
written by P. Reegen. This program checks one frequency after
the other for being a linear combination of previously examined frequencies. If
this attempt fails, the corresponding frequency is considered ``genuine''. Only
genuine frequencies are used to form linear combinations subsequently
\citep[see][for a detailed description]{reegenManual}. Applying this package to the
investigated star, we find only five genuine frequencies. 
All the other frequencies can be explained as simple linear combination of these
terms. We varied the input parameters of Combine to verify the
robustness of the combination calculation. 
The genuine and the combination frequencies are identified in the
last column of Table~\ref{tab1} and in the periodograms of panels  c) to f) of Fig.~\ref{fig2}.

\begin{table*}
 \caption{Pulsation frequencies for CoRoT\,102699796. Columns (1)-(4)
   show frequencies (in c/d and $\mu$Hz), amplitudes and S/N derived
   with $period04$, respectively. Columns (5)-(7) present frequencies,
 amplitudes and $sig$ values obtained with $sigSpec$,
 respectively. Column (8) show the uncertainty on the frequencies
 (after \citet{kallinger08}), whereas column(9) exhibit their
 indentification.}
 \begin{center}
 \begin{tabular}{ccccccccc} 
 \hline
 \hline
 \noalign{\smallskip}
 F$_{P04}$   & F$_{P04}$   & A$_{P04}$  &  S/N$_{P04}$ &  F$_{SigSpec}$   & A$_{SigSpec}$  &  Sig$_{SigSpec}$  &  $\sigma_f$ & Identification     \\
      (c/d)     & $\mu$Hz             &      (ppt)                  &      &           (c/d)                   &      (ppt)        &      &      (c/d)                  &  \\          
       (1)      &    (2)              & (3)     &    (4)         & (5)                 &    (6)  & (7)      &    (8) & (9)    \\
 \noalign{\smallskip}
 \hline
 \noalign{\smallskip}
     6.5584 &    75.91  & 3.47  &   116.5 &  6.5584 &         3.47  &    1585.4  &     0.0002  &              f1       \\                 
     3.3180 &    38.40  & 1.40  &    37.9 &  3.3180 &         1.40  &     497.5  &     0.0003  &              f2       \\        
     6.5069 &    75.31  & 0.96  &    32.1 &  6.5069 &         0.96  &     273.4  &     0.0005  &              f3       \\        
     6.7929 &    78.62  & 0.76  &    25.2 &  6.7929 &         0.76  &     186.6  &     0.0006  &              f4       \\        
     3.2405 &    37.51  & 0.66  &    17.6 &  3.2406 &         0.66  &     147.1  &     0.0006  &              f1$-$f2    \\
     3.5822 &    41.46  & 0.23  &     6.5 &  3.5822 &         0.23  &      19.2  &     0.0017  &              f5       \\        
     3.6603 &    42.36  & 0.18  &     5.2 &  3.6603 &         0.18  &      12.4  &     0.0022  &              2f2$-$f1   \\
   13.1164  &   151.81  & 0.15  &     4.9 & 13.1162 &         0.16  &       8.8  &     0.0026  &              2f1      \\
   13.0645  &   151.21  & 0.14  &     4.7 & 13.0646 &         0.14  &       7.7  &     0.0027  &              f1$+$f3    \\
     6.6361 &    76.81  & 0.13  &     4.3 &  6.6366 &         0.13  &       7.0  &     0.0029  &              2f2      \\
     9.8762 &   114.31  & 0.13  &     4.1 &  9.8761 &         0.13  &       6.6  &     0.0030  &              f1$+$f2    \\
\noalign{\smallskip}  
 \hline
 \end{tabular}
 \end{center}
 \label{tab1}
\end{table*}

\section[]{Spectroscopic observations}

Intermediate-resolution spectroscopy of CoRoT\,102699796 was carried out
with \emph{F}ibre \emph{L}arge \emph{A}rea
\emph{M}ulti-\emph{E}lement \emph{S}pectrograph
\emph{(FLAMES)}, \citep[][]{pasquini02} mounted at the Nasmyth\,A platform of
the VLT 8.2-m unit telescope\,2. The low-resolution gratings \emph{LR02}
and \emph{LR06} were used, leading to a resolving power of $R \approx 6400$ 
and $R \approx 8600$, respectively. The spectral range coverage is about
600\,{\AA}~($3960-4560$\,{\AA}) for the blue grating and about 700\,{\AA}~($6440-7160$\,{\AA})
for the red grating. 

We acquired four exposures with the \emph{LR02} grating,
two during the night of 2009, February 09 and two others during the night 2009,
March 22. In order to increase the signal-to-noise ratio of the final 
spectrum, we corrected each single spectrum to the velocity restframe, we 
verified that no residual lines shift is present at our resolution (i. e., due to 
a hidden companion), and finally we combined all of them in one spectrum with a
total exposure time of 2160 sec, and a SNR of $\sim$\,60.

Only one exposure has been obtained using the \emph{LR06} grating, precisely 
during the night 2009, February 09, with an exposure time of 2400 sec and with
a SNR of $\approx$\,40.

For the following analysis we used the 1-D, wavelength-calibrated spectra as 
reduced by the dedicated Giraffe pipeline\footnote{BLDRS v0.5.3, written at the 
Geneva Observatory, see http://girbldrs.sourceforge.net}.

\begin{table*}
 \caption{Neutral and first ionization state iron lines identified in the range 4000 - 4500 {\AA} for CoRoT\,102699796,
used for effective temperature and iron abundance calculation. For each line we report wavelength and oscillator strength.}
 \begin{center}
 \begin{tabular}{ccccccccc}
 \hline
 \hline
 \noalign{\smallskip}
$\lambda$ & $\log gf$ & El & $\lambda$ & $\log gf$ & El & $\lambda$ & $\log gf$ & El \\  \noalign{\smallskip}
 \hline
 \noalign{\smallskip}
4001.661 & -1.880 & FeI & 4143.415 & -0.204 & FeI & 4250.119 & -0.405 & FeI \\
4002.083 & -3.471 & FeII& 4143.868 & -0.450 & FeI & 4250.787 & -0.710 & FeI \\
4005.242 & -0.610 & FeI & 4147.669 & -2.104 & FeI & 4258.154 & -3.400 & FeII\\
4007.272 & -1.300 & FeI & 4153.900 & -0.270 & FeI & 4260.474 & -0.020 & FeI \\
4009.713 & -1.200 & FeI & 4154.499 & -0.688 & FeI & 4271.153 & -0.349 & FeI \\
4014.531 & -0.200 & FeI & 4154.805 & -0.370 & FeI & 4271.760 & -0.164 & FeI \\
4017.148 & -0.920 & FeI & 4156.799 & -0.620 & FeI & 4273.326 & -3.258 & FeII\\
4021.866 & -0.660 & FeI & 4157.780 & -0.403 & FeI & 4278.159 & -3.816 & FeII\\
4024.725 & -0.710 & FeI & 4158.792 & -0.670 & FeI & 4282.403 & -0.810 & FeI \\
4030.488 & -0.555 & FeI & 4170.901 & -1.100 & FeI & 4294.125 & -1.110 & FeI \\
4032.627 & -2.440 & FeI & 4173.461 & -2.180 & FeII& 4296.572 & -3.010 & FeII\\
4043.897 & -0.826 & FeI & 4174.913 & -2.969 & FeI & 4299.234 & -0.430 & FeI \\
4044.609 & -1.080 & FeI & 4175.636 & -0.670 & FeI & 4303.176 & -2.490 & FeII\\
4045.594 & -0.896 & FeI & 4176.566 & -0.620 & FeI & 4307.902 & -0.070 & FeI \\
4045.812 &  0.280 & FeI & 4177.692 & -3.753 & FeII& 4314.310 & -3.477 & FeII\\
4062.441 & -0.780 & FeI & 4178.862 & -2.480 & FeII& 4315.085 & -0.970 & FeI \\
4063.276 & -0.748 & FeI & 4180.981 & -1.840 & FeII& 4325.762 & -0.010 & FeI \\
4063.594 &  0.070 & FeI & 4181.755 & -0.180 & FeI & 4351.769 & -2.100 & FeII\\
4066.974 & -0.856 & FeI & 4184.891 & -0.860 & FeI & 4352.735 & -1.260 & FeI \\
4067.271 & -1.419 & FeI & 4187.039 & -0.548 & FeI & 4369.411 & -3.670 & FeII\\
4067.978 & -0.430 & FeI & 4187.795 & -0.554 & FeI & 4369.772 & -0.730 & FeI \\
4070.770 & -0.790 & FeI & 4191.430 & -0.666 & FeI & 4375.930 & -3.031 & FeI \\
4071.738 & -0.022 & FeI & 4195.329 & -0.492 & FeI & 4383.545 &  0.200 & FeI \\
4073.762 & -0.920 & FeI & 4198.247 & -0.456 & FeI & 4384.319 & -3.500 & FeII\\
4074.786 & -0.970 & FeI & 4198.304 & -0.719 & FeI & 4385.387 & -2.570 & FeII\\
4075.954 & -3.380 & FeII& 4199.095 &  0.250 & FeI & 4404.750 & -0.142 & FeI \\
4076.629 & -0.360 & FeI & 4202.029 & -0.708 & FeI & 4413.601 & -3.870 & FeII\\
4078.354 & -1.500 & FeI & 4203.938 & -0.350 & FeI & 4415.122 & -0.615 & FeI \\
4081.567 & -0.200 & FeI & 4210.343 & -0.870 & FeI & 4416.830 & -2.600 & FeII\\
4084.492 & -0.590 & FeI & 4216.183 & -3.356 & FeI & 4427.310 & -3.044 & FeI \\
4085.004 & -1.280 & FeI & 4217.546 & -0.510 & FeI & 4442.339 & -1.255 & FeI \\
4085.303 & -0.710 & FeI & 4219.360 &  0.120 & FeI & 4447.717 & -1.342 & FeI \\
4114.445 & -1.220 & FeI & 4222.213 & -0.967 & FeI & 4459.117 & -1.279 & FeI \\
4118.545 &  0.280 & FeI & 4224.171 & -0.410 & FeI & 4461.653 & -3.210 & FeI \\
4121.802 & -1.300 & FeI & 4225.454 & -0.500 & FeI & 4466.551 & -0.590 & FeI \\
4122.668 & -3.380 & FeII& 4227.427 &  0.230 & FeI & 4472.929 & -3.430 & FeII\\
4127.608 & -0.990 & FeI & 4233.172 & -2.000 & FeII& 4476.019 & -0.570 & FeI \\
4128.748 & -3.760 & FeII& 4233.602 & -0.604 & FeI & 4482.170 & -3.501 & FeI \\
4132.058 & -0.650 & FeI & 4235.936 & -0.341 & FeI & 4482.253 & -1.482 & FeI \\
4132.899 & -0.920 & FeI & 4238.810 & -0.280 & FeI & 4489.183 & -2.970 & FeII\\
4134.677 & -0.490 & FeI & 4245.258 & -1.170 & FeI & 4491.405 & -2.700 & FeII\\
4136.998 & -0.540 & FeI & 4247.425 & -0.230 & FeI & 4494.563 & -1.136 & FeI \\
\noalign{\smallskip}
 \hline
 \end{tabular}
 \end{center}
 \label{tabx}
\end{table*}

\subsection{Spectral type and atmospheric parameters}

\subsubsection{Determination of effective temperature}
\label{teff}
Any attempt devoted to a detailed characterization of the chemical abundance
pattern in stellar atmospheres relies on the accuracy of
effective temperature and surface gravity determination.

In this study, we derived the effective temperature by using the ionization
equilibrium criterion. In practice, we adopted as T$_{\rm eff}$ the value
that give the same iron abundance as computed from a sample of spectral lines
(both neutral and first ionization stage) present in the spectral range between
4000 {\AA} and 4500 {\AA} (see Tab.~\ref{tabx}).

We used the method of the spectral synthesis in order to overpass the strong
blending due to relatively high rotational velocity (see below). To perform these calculations
we used ATLAS9 \citep{kur93} to compute LTE atmospheric models and
SYNTHE \citep{kur81} to reproduce the observed spectrum. Line lists and atomic
parameters used in our modeling are from \citet{kur95} and the subsequent update by 
\citet{castelli04}. From the first steps of our iterative procedure, it appeared 
that the star has a low metal content. Thus, we decided to use Opacity Distribution 
Functions [ODF] computed for sub-solar metallicity.
At the end we estimated T$_{\rm eff}$\,=\,7000~$\pm$200~K, computed for
ODF $[M/H]$\,=\,$-$1. The error on the temperature was estimated to be the
the variation in the parameter that increases the $\chi^2$ by unity.

As a by-product, we obtained an estimate of the projected rotational velocity
$v \sin i$\,=\,50\,$\pm$\,5~km/s and an iron abundance
$[Fe/H]$\,=\,$-$1.1\,$\pm$\,0.2. Unfortunately, we were able to obtain only an upper limit for the
magnesium abundance $[Mg/H]\leq-$2.

\begin{figure}
\includegraphics[width=7cm]{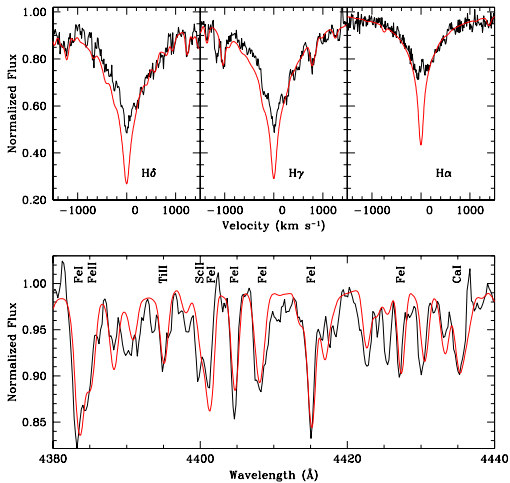}
\caption[]{Top and bottom panels show selected portions of our spectrum,
i.e., the Balmer lines and the 4380\,-\,4440~{\AA} range,respectively.
In both panels the synthetic
spectra is over-imposed (red line).}
\label{spectra}
\end{figure}

In Figs.~\ref{spectra}, we show two portions of our spectral range:
one including the Balmer lines and the other covering the spectral
range 4380\,-\,4440~{\AA}, respectively, with the synthetic
spectra superimposed.

It is worth mentioning that the metallicity measured for
CoRoT\,102699796 is rather low with respect to that expected on the
basis of recent estimates of the Galactic metallicity gradient
\citep[see, e.g.][and references therein]{pedicelli},
i.e. [Fe/H]$\sim$-0.5. However, as discussed in Sect.\ref{membership}, 
it is possible that the region where CoRoT\,102699796 formed,
is associated with the young open cluster Dolidze 25 and with the star
forming complex that include other HII regions (such as Sh 2-284/5),
which is known to host stars as metal poor as [Fe/H]$\sim$-0.8
\citep[][]{lennon}, a 
value close to that measured for
our target star. The reason of such a low metallicity is not
clear, and of course needs to be investigated in detail, however this
is beyond the scope of the present paper.

\subsubsection{Determination of surface gravity and luminosity}
\label{grav}
For early F-type stars, one of the method commonly used in the literature for
the determination of the luminosity class of stars is through the
strength of the ratios between the spectral blend at
$\lambda \lambda$ 4172-4179 {\AA}, mostly Fe{\sc ii} and Ti{\sc ii} lines,
and that at $\lambda$4271 {\AA}, mostly composed of Fe{\sc i} lines
\citep{gray89}.

We computed the $\lambda \lambda$ 4172-4179/4271 ratio which resulted
in 1.60\,$\pm$\,0.15, roughly corresponding to a luminosity class of V \citep{gray89}.

To derive the surface gravity of our target,
we computed the theoretical behavior of the above ratio as a function of $\log g$.
After having fixed T$_{\rm eff}$ to the value found in Sect.~\ref{teff},
we computed a grid of ATLAS9 atmospheric models with gravities spanning the range
between 3.0 and 4.5 dex. The theoretical equivalent widths and consequently
the ratio $\lambda \lambda$ 4172-4179/4271, have been evaluated using the 
code XLINOP \citep{kur81}. By using this curve, we converted our measured
ratio and its associated error in a measurement of gravity and of its uncertainty,
obtaining: $\log g$\,=\,3.8\,$\pm$\,0.4. 

Using the above results, we conclude that CoRoT\,102699796  
has spectral type F1~V.

The atmospheric parameters derived here can be used to estimate the
stellar luminosity. To this aim, we use the
approach by \citet{balona11}. Starting from the
Stefan-Boltzmann law L=L(T$_{\rm eff}$,R) where R is the radius 
and all the quantities are in solar units, we
can eliminate the radius from the definition of gravity g=g(R,M),
where M is the mass, obtaining a relation L=L(T$_{\rm
  eff}$,M,g). Finally, the mass can be eliminated using the
empirical mass-luminosity relation by \citet{malkov07}: $\log
L/L_\odot$=$\log(M/M_{\odot}) = 0.00834+0.213 \log(L/L_{\odot})+0.0107
\log (L/L_{\odot})^2$, thus 
deriving the desired L=L(T$_{\rm eff}$,g) relation.

By using this relation we obtain log(L/L$_{\odot})$=1.33$\pm$0.3 for
CoRoT\,102699796. The relevant stellar parameters for this star are
shown in Table~\ref{tabParameters}.

In Figs.~\ref{spectra}, we show two portions of our spectral range:
one including the Balmer lines and the other covering the spectral
range 4380\,-\,4440~{\AA}, respectively. The synthetic spectra superposed
has been calculated using the atmospheric parameters derived as described 
before and summarized in Tab.~\ref{tabParameters}. 

\begin{table}
 \caption{Physical parameters of $CoRoT$\,102699796 as derived from
   spectroscopy.}
 \begin{center}
 \begin{tabular}{cccccc} 
 \hline
 \hline
\noalign{\smallskip}
   SpT  & T$_{eff}$  & log(g) & [M/H] &   $v$sin$i$ & L \\
             &   (K)        &           &            &     (km/s)   &
             (L$_\odot$) \\
 \noalign{\smallskip}
\hline
\noalign{\smallskip}
  F1V  &  7000$\pm$200   &     $3.8\pm0.4$  & $-$1.1$\pm0.2$ & $50\pm5$ &  21$^{+21}_{-11}$ \\
\noalign{\smallskip}  
\hline
\end{tabular}
 \end{center}
 \label{tabParameters}
\end{table}

\section{The evolutionary status of CoRoT\,102699796}

In this section we deal with the evolutionary status of our
target: is it indeed a young intermediate-mass star (Herbig
Ae type), as we
suspect? We have seen in the previous
section that the Balmer lines in the spectrum appear to be filled. This is the first
necessary clue. The next steps are to investigate if 1) the star shows infrared
excess, and 2) it is possible to associate it with a known star forming
region. 

\subsection{The SED of CoRoT \,102699796}
 \label{sed}
\begin{figure}
\resizebox{\hsize}{!}{\includegraphics[]{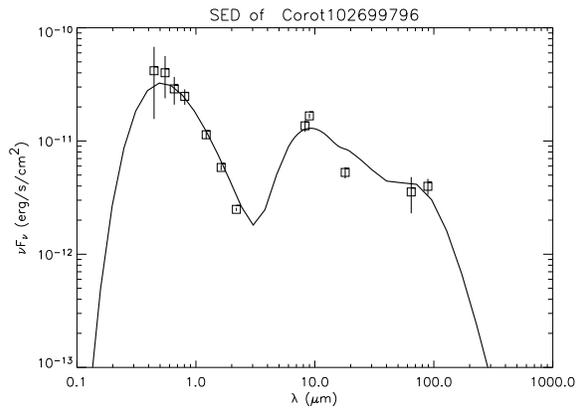}}
\caption{The dereddened SED of CoRoT\,102699796 is shown with open squares.
The continuum line is the total flux given by the model star+disk with the 
parameters presented in Table \ref{tab:disk}. }
\label{img:sed}
\end{figure}

We start by building the Spectral Energy Distribution (SED)
for CoRoT\,102699796.
 A set of photometric data was collected from the literature.
 Observations in the $B,V, R$ and $I$ Johnson bands are present
 in the CoRoT EXO-dat catalogue  \citep{deleuil2009}.
 The NIR data in the $J, H,$ and $K$ bands were taken from the 
 2MASS catalog \citep{2mass,cutri2003}. Four measurements
in the mid- and far-infrared were collected from the 
AKARI Infrared Astronomical Satellite catalogues
IRC \citep[mid-IR all-sky Survey, ][]{ishihara2010} and FIS 
\citep[Far-IR all-sky Survey, ][]{yamamura2010}.
The data were acquired by the Akari-satellite in four bands centered on  
 9, 18, 65 and 90 $\mu$m.  The position accuracy in the first two bands is estimated 
 to be better than 2$"$. At larger wavelengths the pixel size is of 26.8$"$
and the average position disagreement between the AKARI-FIS catalogue and 
the Simbad catalogue is of $\sim$6.5 arcsec. The source identified in
the AKARI-FIS catalogue is at a distance of $\sim$23 arcsec from the 
coordinates given in the CoRoT catalogue for  CoRoT\,102699796. Given 
the pixel sizes of the AKARI camera and 
 the absence of bright sources in a radius of 23 arcsec around CoRoT\,102699796,
  we can state that the observed FIR flux comes from this star.
Additional data in the mid-IR can be retrieved from the database of
the mission Midcourse Space Experiment (MSX) \citep[][]{egan03}. In
particular, the target star was detected in the $A$ band of the satellite
(centered at 8.28 $\mu$), which is close in wavelength to the AKARI $S9W$ band. 
Note that these two independent measures are in
perfect agreement one each other (see Fig.~\ref{img:sed}).

  The magnitudes of the star were dereddened by using the relations 
  given in \citet{cardelli89} between the A$_{\rm v}$ and the other photometric bands.
  In order to estimate  A$_v$, 
  we compared the $B-V$, $V-I$ and $V-R$ colours with the ones 
  obtained using the stellar models of \citet{castelli2004} for the temperature
  and the metallicity of the star. The E(B-V), E(V-R) and E(V-I) 
  so far derived were then converted to $A_{\rm v}$ using the transformations
  of \citet{cardelli89} and averaged. The estimated extinction  
   is $A_{\rm v}=0.93\pm0.06$ mag.

   The  dereddened SED  was compared to  theoretical 
   models of star+disk computed using the software $cgplus$,
    a modeling program, written by C.P. Dullemond, 
     for dusty circumstellar disks, based on the models
    of \citep{chiang1997} and \citep{dullemond2001}. 
   The basic idea is to model the disk around Herbig Ae/Be stars 
   with a passive, irradiated disk with an inner hole.
   In our case the model which best fit the SED is the one with the parameters
   reported in Table \ref{tab:disk}. The SED with the model fit is shown
   in Fig. \ref{img:sed}. We note that the luminosity obtained from
   the integration of the star's SED is in excellent agreement with
   that estimated from the spectroscopy, as reported in
   Tab.~\ref{tabParameters} (see next section for a discussion about
   the distance of the target). \\
 The shape of the SED is typical of transitional circumstellar disks \citep{calvet2005}.
  These usually  have an inner region which is depleted of gas and dust and an outer region 
which is optically thick. The strong flux rise around 10 $\mu$m is presumably due to the frontal
illumination of the wall of the optically thick outer disk. Following the classification scheme 
of \citet{sartori2010} for Herbig Ae/Be stars, 
CoRoT\,102699796 can be identified as a group 2 object, intermediate 
between embedded objects and stars with more evolved dusty disks. 
These group 2 objects can show a doubly peaked SED 
related to a visible central star (first peak) 
surrounded by a significant amount of
 circumstellar cold matter responsible for the second peak in the mid-
 to far-IR band. 
In the group 2 objects 
of \citet{sartori2010} the second peak 
is similar to the first, like in the case of CoRoT\,102699796 (see Fig.~\ref{img:sed}).

\begin{table}
\caption{Parameters derived by the SED fit. }
\label{tab:disk}
\begin{center}
\begin{tabular}{ll}
\hline
\hline
\noalign{\smallskip}
Parameter & value     \\ 
\noalign{\smallskip}
\hline
\noalign{\smallskip}
 Distance &  4000    pc       \\
Inclination  & 60.0$^{\rm o}$\\
Star mass & 2.0 M$_\odot$  \\
Star luminosity &20.0 L$_\odot$\\
Inner disk-radius & 4.0 AU\\
Outer disk-radius & 400 AU\\
Disk mass & 0.06 M$_{\odot}$ \\
Inner disk-temperature & 450 K\\
 Surface density exponent& $-$2.0\\ 
\noalign{\smallskip}
\hline
\end{tabular}
\end{center}
\end{table}

\section{Membership}
\label{membership}
\begin{figure*}
\begin{center}
\includegraphics[width=130.mm]{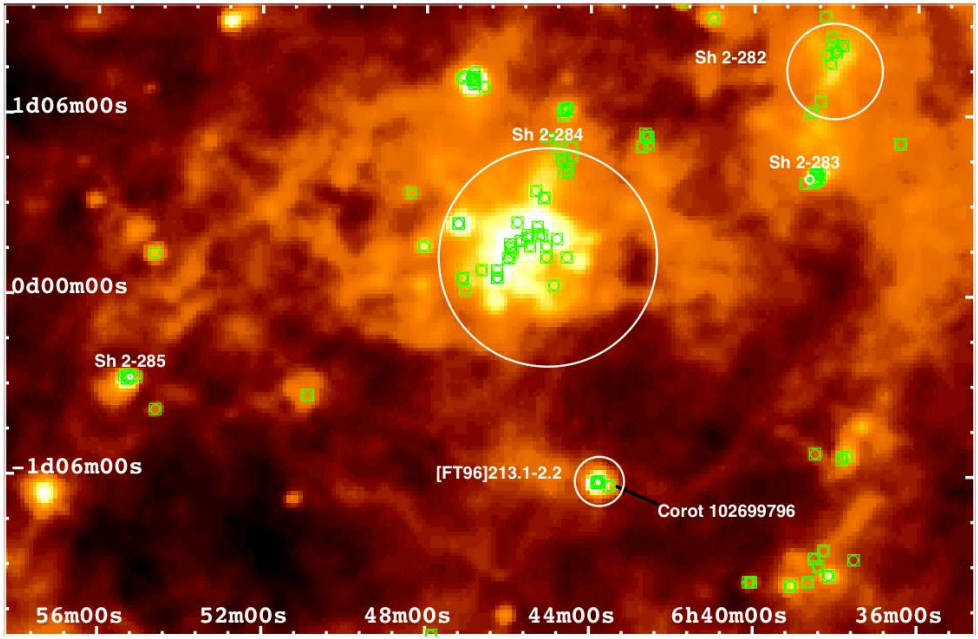}
\caption{Composition of three IRAS image  at 25, 60 and 100 $\mu$m. 
White open circles show the HII regions according to the \citet{sharpless1959} 
catalogue, while green circle-square symbols show young stellar objects
reported by \citet{avedisova2002}. The target star CoRoT\,102699796 is 
inside the circle of [FT96]213.1-2.2. }
\label{img:iras}
\end{center}
\end{figure*}

CoRoT\,102699796 is placed at an angular distance of 1.65 arcmin   
from  the center of the HII region     
[FT96]213.1-2.2 \citep{fich1996}  which extends  over 9 arcmin
 in the Monoceros constellation.  Moreover 
\citet{avedisova2002} lists 7 young objects 
close ($\le$ 4 arcmin) to CoRoT\,102699796. 
We believe, indeed, that the CoRoT object, together 
with the 7 young objects, belongs to the
[FT96]213.1-2.2 HII region, which is in turn 
associated with a big star formation structure
composed by several HII regions, such as Sh 2-284, Sh 2-283 and 
Sh 2-285 \citep{russeil2007}. 
The distance estimate of this  
structure is still controversial. \citet{russeil2007} derived, using 
also data from literature,  a distance of 7.89 kpc. 
Nonetheless one of the biggest part of the star formation structure, the HII region Sh 2-284,
was found to be at a distance of 4 kpc by \citet{cusano},  in agreement
within the errors with the distance of 3.6 kpc given by \citet{delgado}. 
Furthermore, \citet{rolleston1994} also found that 
two B0V stars in Sh 2-285 are at a  distance of 4.3 kpc.
These results combined together,  place the 
star formation region at a distance  of 4-4.5 kpc, closer then the
estimate of \citet{russeil2007}. The luminosity of CoRoT\,102699796 
derived by the frequency analysis (as estimated in Sect~\ref{sectTEO}), by the spectroscopic
observations and by the SED fitting (see Section \ref{sed}), is also
compatible with a distance of the object of 4 kpc.
The three colour image shown in Figure \ref{img:iras}, obtained by combining IRAS images at
25, 60 and 100 $\mu$m,  illustrates the different components 
of this large star forming complex. 
The white open circles are HII regions from the \citet{sharpless1959} 
catalogue, while the green circle-square symbols show 
young stellar objects from  \citet{avedisova2002}. CoRoT\,102699796 is 
inside the circle of [FT96]213.1-2.2. 

In the spectroscopic analysis, CoRoT\,102699796 was found to have a 
metallicity very similar to that of Sh 2-284,  supporting the assumption of 
a common origin of the [FT96]213.1-2.2   and  Sh 2-284 HII regions. 
If confirmed, this complex has one
of the lowest metallicity galactic known SFRs.

\section{Interpretation of the observed oscillation frequencies}

\subsection{Calculation of asteroseismic models}

The model analysis performed in this paper is described in
  detail in \citet{dicriscienzo2008}, and was originally suggested by
  \citet{guenther2004}.  In summary, given the observed N frequencies
  of the investigated star, we look for models having
  $\chi$$^{2}$$\le$1 within our grid of oscillation spectra, with the
  following definition of $\chi$$^{2}$
\begin{equation} 
\chi^2=\frac{1}{N} \sum_{i=1}^N
\frac{(f_{obs,i}-f_{mod,i})^2}{\sigma_{obs,i}^2+\sigma_{mod,i}^2} \label{eq1}
\end{equation}
where $f_{obs,i}$ and $f_{th,i}$ are the observed and
the model frequency for the ith mode, respectively, characterized by an
observational $\sigma_{obs,i}$ and a theoretical $\sigma_{th,i}$
uncertainty. In general, the search for the best-fitting model
(corresponding to the oscillation spectrum that minimizes the
$\chi$$^{2}$ value) is concentrated in the subgrid of models with
T$_{\rm eff}$ and L/L$_{\odot}$ within the error box in T$_{\rm eff}$ and L/L$_{\odot}$ of
the star. If the program does not find any model matching all the
frequencies with $\chi$$^{2}$$\le$1, it discards one by one the
non-matched frequencies, computes the $\chi$$^{2}$ for the
remaining frequencies and searches again for models with
$\chi$$^{2}$$\le$1. Among the resulting models satisfying the condition
$\chi$$^{2}$$\le$1, the model with the lowest value of the $\chi$$^{2}$
is the best-fitting one.
The grid of PMS models where constructed
using the ATON code for stellar evolution \citep[]{ventura98} in the
standard version for asteroseismic applications
\citep[]{dantona2005}. The physics of the models are up-to-date and
described in detail in \citet{dicriscienzo2008}.  The ATON code also
allows to include the effects of rotation, according to the
formulation by \citet{endal1976}, as described in
\citet{mendes1999,landin2006}.  The possibility of accounting for the
rotation effects proves to be extremely useful for this analysis,
since PMS stars have on average high rotational velocities
\citep{bohm1995}. In general, what we may determine is the projected
velocity v$\sin$i, but in this particular case we can derive the true
rotational velocity ($\sim$60km/s) from the inclination of the disk,
assuming the disk is in the stellar rotation plane,
and this can be used to further fix the best fit model.\\
We stress here that the approach by \citet{endal1976} accounts only
for the hydrostatic effects of rotation, neglecting the internal
angular momentum redistribution; in this work we use a rigid body
rotation. The initial angular momentum of the star is provided as a
physical input, and chosen to reproduce the observed
surface angular velocity of the star at a given position in the HR diagram.\\

\begin{figure}
\centering
\includegraphics[width=8.5cm]{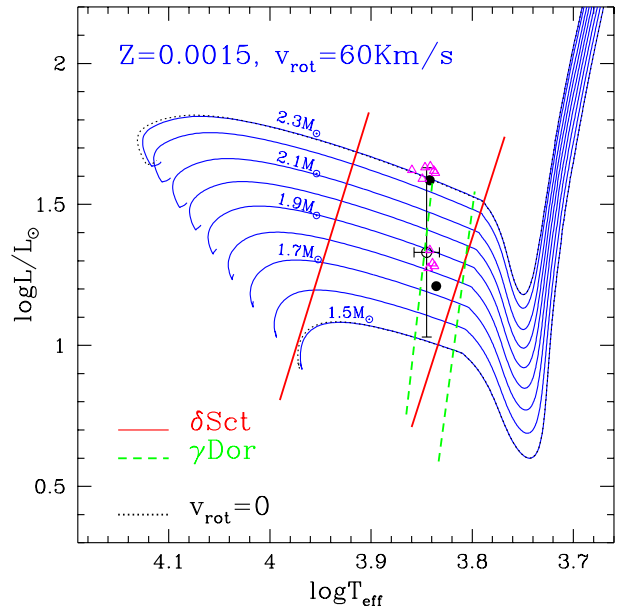}
\vspace{-1.5cm}
\caption{HR diagram showing the spectroscopic position of
  CoRoT\,102699796 (open circle) with the respect to several PMS
  rotational tracks taken from the model grid. The black full circles mark
  the position of the best fit models found comparing observed and
  predicted frequencies with m=0 while magenta triangles are the
  models described in Table 6. 
The red solid lines and the green dashed 
lines show the $\delta$\,Sct and $\gamma$\,Dor instability strips by
\citet{breger98} and \citet{guzik00}, respectively.}
\label{HR}
\end{figure}

Models were computed from 1.5 to 2.3 M$_{\odot}$, because these are
the masses that enter in the empirical box in the HR diagram (see
Fig.~\ref{HR}), with a step of $\Delta$M=0.01 M$_{\odot}$. The
evolution begins on the Hayashi track, and the calculation ends when
the star has reached the MS. We note that neglecting the protostellar
evolution and its effect on the inner structure and pulsation
properties does not change the conclusions of the paper because the
investigated star lie on the horizontal portion of the PMS
evolutionary tracks, subsequent to the thermal relaxing phase
undergone by stars around 2 $M_{\odot}$ in the earliest phases of the
PMS evolution. The step in effective temperature between consecutive
models of
the grid is $\Delta$$T_{\rm eff}$ $\sim$50K.\\
We use the LOSC oscillation code by \citet{scuflaire2008} to calculate
the frequencies of radial and non radial oscillation modes for each
evolutionary structure of the grid.  We limit the spherical
  degree to l$<$4 and contrary to previous works on PMS pulsating
stars we extend our pulsational analysis also to gravity modes since
the oscillation spectra of this star has an anomalous extension to low
frequencies suggesting that at least the lowest ones could be low
order $g$-modes.

\subsection{Comparing observed and predicted frequencies} 

\label{sectTEO}

\begin{figure}
\centering
\includegraphics[width=8.5cm]{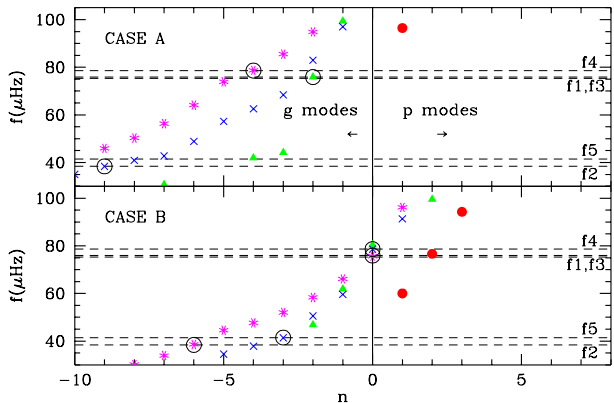}
\vspace{-5cm}
\caption{Comparison between the observed spectrum of star 
CoRoT\,102699796 and the spectrum of the best fit model obtained in the
  two cases indicated in Table \ref{tabellaMOD}. Filled circles
  and triangles, crosses and asterisks indicate theoretical frequencies 
with l=0, 1, 2, 3, respectively. The open circles indicate the matched frequencies}
\label{FREQ}
\end{figure}

Among the observed frequencies reported in Table~\ref{tab1} we try to
fit those which are not consistent with a linear combination of the
others, namely f1 to f5, in the order adopted in Table~\ref{tab1}.
 We also start  excluding from the analysis $f_3$ which is too similar to
  $f$$_{1}$ and we suggest that these two frequencies are an example
  of frequency pairs (separations lower than 0.7$\mu$Hz or 0.06 c/d)
  present in the power spectra of the majority of the well-studied
  $\delta$ Sct stars, as noted by \citet{breger2002}. Since the actual
  cause for this
  phenomenon is still unknown we prefer to remove $f_{3}$ from our fitting procedure.\\
  In this way we have four observed frequencies to be fitted with
  computed ones in order to find the best fit model through the
  $\chi$$^{2}$ analysis.\\
  
 We assume as uncertainties on observed frequencies  in
    eq.~\ref{eq1}  values which are  3-$\sigma$ the observational errors 
given in Section 2.1  and consider conservatively theoretical errors 
of the same order of magnitude  of the observational ones, taking into 
account all the intrinsic numerical uncertainties of models 
\citep[see][for a detailed quantitative description]{moya2007}.\\

  As a result we find that the best fit model corresponds to a very young star
    (t$\sim$3 Myr, with no convective core yet) with M=1.74
    M$_{\odot}$, T$_{\rm teff}$=6850K (CASE A in Table
    \ref{tabellaMOD}). We note that the $\chi$$^{2}$ reported in
    table is calculated on the basis of the first three
    frequencies because unfortunately this model does not match 
    frequency f5, which however is the one with the lowest S/N ratio (see
    Tab.~\ref{tab1}).  Moreover, if we allow $\chi$$^{2}$ to be $>1$, 
    we find that a
    much more massive model (described in Table~\ref{tabellaMOD} as
    CASE B) minimizes $\chi$$^{2}$. According to this model, only two
    frequencies are pure g-mode of low order.

We remark that in the analysis described above the
 rotation of the star (see Sect. 3) was considered only in the
 computation of the evolutionary model.
For completeness we recall
that if $f$ denotes the frequency in the absence of rotation, a slow
solid rotation with angular velocity $\Omega$ slightly alters the
frequency with l$>$0 in the following way:

\begin{equation}
f_{l,n,m}=f_{l,n}+m\beta_{l,n}\Omega \label{eq2}\\
\end{equation}

where $\beta$ is a quantity described in detail in
\citet{dicriscienzo2008} and calculated by the adiabatic oscillation code for each
pair (l,n). 
We then relaxed the
 previous constraint used to find the best fit model by adding  the
 multiplet 
(triplet and quintuplet, etc. according to eq.~\ref{eq2})
frequencies to our  predictions, and we compared again theory 
and observations using the same technique as above (but this time
trying to fit also $f_3$). As a result, the number of best-fitting models grows in number, as
shown in Tab.~\ref{tabellaMOD2}, where we list all the models between
$\chi^{2}_{min}$ and $\chi ^{2}_{min}+1$ (i.e. within 1 $\sigma$). However,
when we report these models in the HR diagram (empty triangles in
Fig. \ref{HR}), we find out the very interesting result that these
models cluster around or close to the two best fit models previously
found, namely CASE A and B. This occurrence suggests  that models
consistently predict two possible solutions for the position of the
star in the HR diagram: a) with  LogL/L$_{\odot} \sim $ 1.2$\div$1.3
and  M/M$_{\odot} \sim $ 1.8; b) with LogL/L$_{\odot} \sim $ 1.6
and  M/M$_{\odot} \sim $ 2.3. 
To decide which ``class'' of models has to be preferred, we note that
the ``class a)'' are the favoured models because: i) they produce the 
lowest values for the $\chi$$^{2}$ and ii) their luminosities are in better agreement with
the values inferred from spectroscopy and from the integration of the SED  (although at least formally also the
luminosity of class b) models are in agreement with the
spectroscopic value at 1 $\sigma$). On the basis of the above
considerations, our 
best-fit  model is the one that minimize the $\chi$$^{2}$ (model a1 in Tab.~\ref{tabellaMOD2}), with 
mass, effective temperature and luminosity of  1.84  M/M$_{\odot}$,
6900 K,  and 1.29  LogL/L$_{\odot}$, respectively. Similarly to CASE
A, it is a young model ($\sim$2.5 Myr) still without a convective core.
It is important to note that the observed frequency
spectrum can be interpreted in terms of $g$-modes with low-moderate
n-value for class a) models, whereas class b) models are characterized
by both $p$- and $g$-modes.

\begin{table*}
\centering
\caption{Mode identification considering all frequencies with m=0}
\begin{tabular}{lcccc|cccc}
\hline\hline  
\noalign{\smallskip}
&&&&&$f$$_{obs,1}$&$f$$_{obs,2}$&$f$$_{obs,4}$&$f$$_{obs,5}$\\
&&&&&75.91&38.40&78.62&41.46\\
&&&&&($\mu$Hz)&($\mu$Hz)&$(\mu$Hz)&($\mu$Hz)\\
\noalign{\smallskip}
\hline\hline  
\noalign{\smallskip}
 CASE &$\chi$$^{2}$& M/M$_{\odot}$& T$_{\rm eff}$(K) & LogL/L$_{\odot}$ &$f$$_{th,1}$(l,n)&$f$$_{th,2}$(l,n)&$f$$_{th,4}$(l,n)&$f$$_{th,5}$(l,n)\\
\hline
A&0.83& 1.74 &  6850 & 1.21& 75.89(1,-2)&38.36(2,-9)&78.67(3,-4)&41.98\footnote{This frequency is not considered in the computation of $\chi$$^{2}$ (see text)}(1,-4)\\ 
B&4.92 & 2.29 & 6950  & 1.59&75.66(3,0)&38.57(3,-6)&78.59(2,0)&41.33(2,-3)\\
\noalign{\smallskip}
\hline
\end{tabular}
\label{tabellaMOD}
\end{table*}

These results, if confirmed, would be of crucial importance because it
would be the first time that $g$-modes are detected in a Herbig Ae
pulsating star.  An even more complicate picture emerges if we
consider that CoRoT\,102699796 lies (see Fig. ~\ref{HR}) in the region
of the HR diagram where $\delta$ Sct and $\gamma$ Dor instability
strips intersect \citep[at least for solar metallicity, see
e.g.][]{handler}. We recall that the frequency spectrum of $\gamma$
Dor stars is interpreted in terms of high radial order $g$-modes. Our
findings suggest that CoRoT\,102699796 shows a frequency spectrum with
intermediate properties between $\delta$ Sct and $\gamma$ Dor,
characterized by frequencies lower and higher than those typical of
$\delta$ Sct and $\gamma$ Dor variables, respectively.  In fact,
according to our interpretation this star is a pre-main sequence
pulsator which shows $g$-modes of low-moderate radial number and order in its
frequency spectrum (or both $p$- and $g$-modes, see above).  In this context, it is worth noticing that hybrid
$\delta$ Sct-$\gamma$ Dor pulsators were already found among more
evolved stars by several authors\citep[see,
e.g.][]{henry05,king06,rowe06,handler08}. Moreover, the recent data
obtained with the KEPLER satellite seem to show that the hybrid
phenomenon is more
common than expected \citep[][]{grigahcene10,catanzaro10}. \\
However, prior to arrive to firm conclusions concerning the nature of
CoRoT\,102699796, a more detailed theoretical study is necessary by
using a non-adiabatic oscillation code for a stability analysis of the
selected modes.
Furthermore the main feature of non-adiabatic codes as 
MAD \citep[][]{dupret03} is their ability to determine accurately 
the amplitudes and  phases of variation of different physical 
quantities in the very outer layers of the stars. In particular the 
normalized amplitude and phase of effective temperature variations 
can be determined, significantly improving the accuracy of the mode 
identification methods through multi-band photometric approaches. 
An important point is that the non-adiabatic predictions are very
sensitive to the modelling of the thin convective envelope of these 
stars and to the treatment of its interaction with pulsation, thus only  
time dependent convective versions of  these codes would be able 
to  reproduce both the blue and the red edge of the instability 
region of each mode.\\

\begin{table*}
\centering
\caption{Mode identification adding in the simulated  spectra multiplet frequencies}
\begin{tabular}{ccccc|ccccc}
\hline\hline  
\noalign{\smallskip}
&& & & & $f_{obs,1}$ & $f_{obs,2}$ & $f_{obs,3}$ & $f_{obs,4}$ & $f_{obs,5}$  \\
&&&&&75.91&38.40&75.61&78.62&41.46\\
&&&&&($\mu$Hz)&($\mu$Hz)&($\mu$Hz)&$(\mu$Hz)&($\mu$Hz)\\
\noalign{\smallskip}
\hline\hline  
\noalign{\smallskip}
MODEL &$\chi$$^{2}$ & M/M$_{\odot}$ & T$_{\rm eff}$(K) & LogL/L$_{\odot}$ & l,m,n & l,m,n & l,m,n & l,m,n & l,m,n \\
\hline
a1 &0.28 & 1.84   &   6900 & 1.29 &      3, 1,-3    &    3,1,-8 &  3,2,-5 &     3,-2,-2      &      3,-1,-8  \\
b1 &0.37 & 2.28   &   7050 & 1.59 &      3,-1,0    &    3,1,-7 &  2,-2,0   &     2,-2,0       &      2,2,-5\\
b2 &0.46 & 2.36   &   7000 & 1.63 &      3,-2,1     &    3,2,-7 &  2,0,0    &     2,1,0        &      2,2,3\\
a2 &0.84 & 1.90   &   6950 & 1.34 &      3,3,-5     &    1,0,-3 &  2,1,-2   &     3,-1,-2      &      3,2,-4\\
b3 &0.86 & 2.32   &   7250 & 1.62 &      2,1,0      &    2,0,-3 &  3,2,0    &     2,2,0        &      3,-1,-3\\
a3 &0.88 & 1.83   &   6875 & 1.28 &      3,1,-3     &    1,-1,-3&  2,0,-2   &     3,-2,-2      &      3,2,-15\\
b4 &0.90 & 2.35   &   6850 & 1.61 &      2,1,0      &    2,0,-3 &  3,2,0    &     2,2,0        &      3,2,-6\\
b5 &0.90 & 2.38   &   6950 & 1.64 &      3,-1,1     &    2,0,-3 &  3,2,0    &     2,2,0        &      2,2,-6\\
b6 &1.00 & 2.36   &  6900 & 1.61 &      3,-1,1     &    2,0,-3 &  3,2,0    &     2,2,0        &      2,2,-6\\
a4 &1.14 & 1.81   &  6975 & 1.27 &      2,-1,-2    &    1,-1,-4&  3,-3,-5  &     3,-3,- 2     &      3,3,-19\\
\noalign{\smallskip}
\hline
\end{tabular}
\label{tabellaMOD2}
\end{table*}

\section{Discussion and conclusions}

In this paper we presented a comprehensive study of 
CoRoT\,102699796, a star observed by the CoRoT satellite during its
first Long Run in the anticenter direction. The star was classified as
$\delta$ Sct and shows anomalous Balmer lines at a first analysis of 
mid-resolution spectra obtained with GIRAFFE@VLT, thus being a likely
candidate of the PMS $\delta$ Sct class. 

We analysed the time-series photometry observed by the satellite CoRoT, detecting the presence of
five independent oscillation frequencies in the range 3-6.5 c/d. These
low values are somewhat atypical for PMS and ``normal'' $\delta$ Sct
variables, whose peaks are usually placed around 10-30 c/d. 

The star's physical parameters were estimated on the basis of the
quoted GIRAFFE@VLT mid-resolution spectra, finding that  the star has
spectral type F1V, T$_{\rm eff}$=7000$\pm$200 K, log(g)=$3.8\pm0.4$,
[M/H]=-1.1$\pm0.2$, $v$sin$i$=$50\pm5$ km/s, L/L$_{\odot}$=21$^{+21}_{-11}$. 
Therefore CoRoT\,102699796 is the first intermediate-mass 
PMS pulsating star for which poor metallicity has been reported.

The evolutionary status of the target star has been explored in
detail by constructing the SED, extending from the optical to
the mid-infrared, using ground-based and satellite data. As a result,  we
find that the shape of the SED is typical of transitional
circumstellar disks.  Following the classification scheme 
of \citet{sartori2010} for Herbig Ae/Be stars, 
CoRoT\,102699796 can be identified as a group 2 object, i.e. between
embedded objects and stars with more evolved dusty
disks. Its PMS nature is reinforced by the possible association with the
HII region [FT96]213.1-2.2 \citep{fich1996}.

The pulsation frequencies have been interpreted in the light of the
non-radial pulsation theory, using the LOSC program \citep{scuflaire2008} in conjunction with
static and rotational evolutionary tracks calculated by means of the
ATON code for stellar
evolution \citep[]{ventura98} in the standard version for
asteroseismic applications with low metallicity \citep[]{dantona2005}.
A minimization algorithm
was used to find the model that matches the observed frequencies. As a
result, either including or not the rotation, we find two possible solutions for the position of the
star in the HR diagram: a) with  LogL/L$_{\odot} \sim $ 1.2$\div$1.3
and  M/M$_{\odot} \sim $ 1.8; b) with LogL/L$_{\odot} \sim $ 1.6
and  M/M$_{\odot} \sim $ 2.3. 
Models with lower mass and luminosity are preferred because: i) they
minimize the $\chi$$^{2}$ and ii) their luminosities are in better agreement with
the values inferred from the spectroscopically estimated logg and from
the integration of the star's SED. Finally, the model named a1 in
Tab.~\ref{tabellaMOD2} is our best-fit model (minimum $\chi$$^{2}$ using
all the five significant frequencies), having  
mass, effective temperature, and luminosity equal to  1.84  M/M$_{\odot}$,
6900 K,  and 1.29  LogL/L$_{\odot}$, respectively. The position in the
HR diagram of this model is in excellent agreement with the
spectroscopic one. The model is very young 
(age$\sim $2.5Myr) without a convective core, in agreement with the shape 
of the SED that shows a significant FIR excess. 
The theory-observation match is based on the interpretation of
frequencies $f_1 \div f_5$ as $g$-modes of low-moderate n-value. 
To our knowledge, this is a
new result and the first time that such modes are observed in 
PMS intermediate-mass pulsating stars. However, this is not completely surprising since it is also the first
time that a so low frequency range is observed  in a pulsating Herbig
Ae star. In this context, it is also remarkable that 
CoRoT\,102699796 lies in the region of HR diagram where 
$\delta$ Sct and $\gamma$ Dor instability strips intersect. Hence we
are probably dealing with a star showing intermediate (hybrid) characteristics
in between the two variability classes. However, we note that if
  the solution predicted by class a) models applied, the star could not
  properly be defined as an hybrid because only $g$-modes would be
  present. In this case it would be better defined as a PMS anomalous
  $\gamma$ Dor star.

\section*{Acknowledgments}

We warmly thank our referee, Torsten B\"{o}hm, for his pertinent and 
useful comments that helped us to improve the manuscript.
It is a pleasure to thank Ennio Poretti for many helpful
discussions. 
This work was supported by the Italian ESS project, contract ASI/INAF
I/015/07/0, WP 03170 and by the European Helio- and Asteroseismology
Network (HELAS), a major international collaboration funded by the European 
Commission's Sixth Framework Programme. \\
This research has made use of the SIMBAD database, operated at CDS, Strasbourg, 
France. This publication makes use of data products from the Two Micron All Sky 
Survey, which is a joint project of the University of Massachusetts and the 
Infrared Processing and Analysis Center/California Institute of Technology, funded 
by the National Aeronautics and Space Administration and the National Science 
Foundation. \\ 
Piet Reegen, author of several
  analysis packages used in this work, has passed away a few months
  ago. We wish to give our deepest sympathy to his relatives and
  friends.

\end{document}